\documentclass[11pt]{article}
\newcommand{\Comment}[1]{{}}
\usepackage{amsfonts,amsthm,amsmath,amssymb,slashed}
\usepackage[textwidth = 430 pt, textheight = 630 pt]{geometry}
\usepackage{color}

\Comment{\usepackage{color}
\definecolor{MyDarkBlue}{rgb}{0.15,0.15,0.45}
\usepackage[linktocpage=true]{hyperref}
\hypersetup{
colorlinks=true,
citecolor=MyDarkBlue,
linkcolor=MyDarkBlue,
urlcolor=MyDarkBlue,
pdfauthor={Jeff Murugan and  Horatiu Nastase},
pdftitle={Duality of topological insulators and superconductors and particle-vortex duality},
pdfsubject={hep-th}
}

\usepackage[numbers,sort&compress]{natbib}
\usepackage{hypernat}}
\usepackage{graphicx}
\usepackage{cite}

\newcommand\ignore[1]{}
\def\one{{\,\hbox{1\kern-.8mm l}}}

\def\d{\partial}

\newcommand{\Cset}{{\,\,{{{^{_{\pmb{\mid}}}}\kern-.45em{\mathrm C}}}}}

\newcommand{\be}{\begin{equation}}
\newcommand{\bea}{\begin{eqnarray}}

\newcommand{\ee}{\end{equation}}
\newcommand{\eea}{\end{eqnarray}}

\begin{document}

\renewcommand{\thefootnote}{\fnsymbol{footnote}}

\makeatletter
\@addtoreset{equation}{section}
\makeatother
\renewcommand{\theequation}{\thesection.\arabic{equation}}

\rightline{}
\rightline{}

\begin{flushright}
QGASLAB-16-01
\end{flushright}

\vspace{10pt}


\begin{center}
{\LARGE \bf{\sc Particle-vortex duality in topological insulators and superconductors}}
\end{center} 
 \vspace{1truecm}
\thispagestyle{empty} \centerline{
{\large \bf {\sc Jeff Murugan${}^{a,b}$}}\footnote{E-mail address: \Comment{\href{mailto:jeff.murugan@uct.ac.za}}{\tt jeff.murugan@uct.ac.za}}
{\bf{\sc and}}
{\large \bf {\sc Horatiu Nastase${}^{c}$}}\footnote{E-mail address: \Comment{\href{mailto:nastase@ift.unesp.br}}{\tt nastase@ift.unesp.br}}
                                                          }

\vspace{.5cm}

\centerline{{\it ${}^a$The Laboratory for Quantum Gravity \& Strings, }} 
\centerline{{\it Department of Mathematics and Applied Mathematics, }} 
\centerline{{\it University of Cape Town,}} 
\centerline{{\it Private Bag, Rondebosch, 7700, South Africa}}
\vspace{.3cm}
\centerline{{\it ${}^b$School of Natural Sciences,}}
\centerline{{\it Institute for Advanced Study, Olden Lane, Princeton,}}
\centerline{{\it NJ 08540, USA}}
\vspace{.3cm}
\centerline{{\it ${}^c$Instituto de F\'{i}sica Te\'{o}rica, UNESP-Universidade Estadual Paulista}} 
\centerline{{\it R. Dr. Bento T. Ferraz 271, Bl. II, Sao Paulo 01140-070, SP, Brazil}}

\vspace{1truecm}

\thispagestyle{empty}

\centerline{\sc Abstract}

\vspace{.4truecm}

\begin{center}
\begin{minipage}[c]{380pt}
{\noindent We investigate the origins and implications of the duality between topological insulators and topological superconductors in three and four spacetime dimensions. In the latter, the duality transformation can be made at the level of the path integral in the standard way, while in three dimensions, it takes the form of ``self-duality in odd dimensions". In this sense, it is closely related to the particle-vortex duality of planar systems. In particular, we use this to elaborate on Son's conjecture that a three dimensional Dirac fermion that can be thought of as the surface mode of a four dimensional topological insulator
is dual to a composite fermion. 
}
\end{minipage}
\end{center}

\vspace{.5cm}

\setcounter{page}{0}
\setcounter{tocdepth}{2}

\newpage

\renewcommand{\thefootnote}{\arabic{footnote}}
\setcounter{footnote}{0}

\linespread{1.1}
\parskip 4pt



\section{Introduction}
The study of the quantum aspects of matter that has facinated physicists for the better part of a century has taken on new impetus with the 
discovery of so-called {\it topological quantum matter}, whose properties are not captured within the Landau symmetry breaking paradigm. 
Chief among these properties is the emergence of a new kind of topological order that encodes patterns of quantum entanglement. Thought of in 
these broad terms then, topological quantum matter can be classified into two categories. {\it Topologically ordered states} encode long-range 
quantum entanglement and contain non-trivial boundary states \cite{Wen:1989iv}. These two facets alone have earned topologically ordered 
states a coveted position at the forefront of the quest to build a robust, fault tolerant quantum computer \cite{Kitaev:1997wr}.

{\it Topological insulators}, in contrast, encode only short-range entanglement. Like topologically ordered matter, they are also characterised by 
non-trivial, gapless boundary states. Remarkably however, in this case, these conducting surface states are protected by various rotational and 
time-reversal symmetries, putting topological insulators into the new category of symmetry protected topological (SPT) quantum 
matter \cite{Gu:2009dr}, and squarely in the crosshairs of this note. In the seminal work \cite{Qi:2008ew}, an effective action for a 3+1 
dimensional (time reversal invariant) topological insulator was proposed, based on dimensionally reducing an auxillary 4+1 dimensional 
topological insulator. The action takes the form of a theta term which, when supplemented with a standard Maxwell kinetic term for the gauge field, reads
\be
  S_{\rm TI}^{(3+1)}=\int d^{3+1}x \left(-\frac{1}{4 e^2}F_{\mu\nu} F^{\mu\nu}+\frac{\theta(\vec{x},t)}  
  {32\pi^2}\epsilon^{\mu\nu\rho\sigma}F_{\mu\nu}F_{\rho\sigma}\right),\label{topinsact}
\ee
where, as usual, $F_{\mu\nu}=\d_\mu A_\nu-\d_\nu A_\mu$ is the field strength of the $U(1)$ electromagnetic field $A_\mu$.

On the other hand, a {\it topological superconductor} is a superconductor with fully gapped quasi-particle excitations in the bulk - the Cooper 
pairs responsible for superconductivity - but has topologically protected, gapless quasi-particle states on the boundary. The latter are, of course, 
responsible for the surface conduction of the superconductor. In a similar way to the topological insulator, it was recently argued in \cite{Qi:2012cs} 
that the 3+1 dimensional topological superconductor has the effective action which, in the simplest case of two Fermi surfaces and first Chern 
number one, reads (see also the later development in \cite{Braga:2016atp})
\bea
  S_{\rm TSC}^{(3+1)}&=&\int d^{3+1}x\left[\frac{\theta_L-\theta_R}{64\pi^2}\,\epsilon^{\mu\nu\rho\sigma}  
  \,F_{\mu\nu}F_{\rho\sigma}
  -\frac{1}{4e^2}F_{\mu\nu}F^{\mu\nu}+\frac{1}{2}\rho_L(\d_\mu\theta_L-2A_\mu)^2\right.\cr
  &&\left.+\frac{1}{2}\rho_R(\d_\mu\theta_R-2A_\mu)^2+J\cos(\theta_L-\theta_R)\right].  
  \label{topsupercact}
\eea
Here $\theta_{L,R}$ are functions on two Fermi surfaces (left and right, from the way they were obtained by a dimensional reduction of an 4+1 
dimensional interval) associated with the phase of the Cooper pairs; $\rho_{L,R}$ refers to the density of Cooper pairs on the two surfaces 
($\rho\sim |\Phi|^2$); and the cosine term describing a possible Josephson coupling. 

One spatial dimension down, in 2+1 dimensions, it was well known that the topological response of a class of topological insulator is described by 
a Chern-Simons type action 
\be
  S_{\rm TI}^{(2+1)}=\frac{e^2}{\hbar}\frac{C_1}{4\pi}\int d^{2+1}x\, \epsilon^{\mu\nu\rho}\,A_\mu \d_\nu   
  A_\rho,\label{topinscs}
\ee
with $C_1=\frac{1}{2\pi}\int dk_x\, dk_y\, f_{xy}(\vec{k})\,\,\in \mathbb{Z}$, being the first Chern number of the Berry connection $a_i$ of the 
insulator bands. As in the higher dimensional case, this action may be supplemented with the usual Maxwell term for the gauge field dynamics. 

The focus of this article is the origins of the duality relation between topological insulators and the topological superconductors. In three spacetime dimensions, 
we will show that this relationship arises as a consequence of the duality between particle and vortex-like excitations peculiar to the 
plane \cite{Zee:2003mt}. This particle-vortex duality has a long history that goes back to the early work on superconductivity of \cite{Dasgupta:1981zz} 
and later in the study of anyon superconductivity and the fractional quantum Hall effect in \cite{Lee:1989fw} (see also the early work 
in \cite{Marino:1987tk,Marino:1992uu}). While the duality can be defined at the level of the path integral \cite{Burgess:2000kj} and even embedded 
into the gauge/gravity correspondence \cite{Murugan:2014sfa} (see also \cite{Ramos:2005yy,Ramos:2007hk} for another take on a path integral formulation), 
the formulation that will be most useful for our purposes is the transformation that takes ``self-duality in odd dimensions" \cite{Townsend:1983xs}
to a topologically massive theory \cite{Deser:1984kw}. This will furnish the necessary tools we need to understand Son's conjectured equivalence 
between a massless Dirac fermion understood as the boundary mode of a topological insulator, and the composite fermion of an effective low-energy theory 
for the half-filled Landau level of a Fermi liquid \cite{Son:2015xqa} (A related duality is the proposed "3 dimensional bosonization duality" proposed at 
the level of supersymmetric theories in \cite{Aharony:2015mjs} and originally proposed in \cite{Burgess:1994tm,Fradkin:1994tt}.).
It was argued already in \cite{Metlitski:2015eka} that this conjecture should be derivable from a (fermionic) particle-vortex duality, 
but the arguments given there were rather implicit. Here we revisit this issue and show that it can be derived from the particle-vortex duality as formulated 
in \cite{Burgess:2000kj}. 

The paper is organized as follows. In section 2 we make explicit the duality between four dimensional topological insulator and topological superconductor, elaborating on some of its physical consequences. Section 3 is devoted to understanding the duality in 2+1 dimensions. In particular we show how to relate it to odd dimensional self-duality and particle-vortex duality. In section 4 we show that Son's conjecture can be understood as a consequence of this particle-vortex duality and conclude in section 5. 

\section{Four dimensional topological superconductor - topological insulator duality}

To summarise our introductory comments, the actions for a topological insulator and a topological superconductor are given by (\ref{topinsact}) and (\ref{topsupercact}) respectively. 
Let's start by thinking about the topological superconductor. Taking the phases, $\theta_L=$const. and $\theta_R=$const. and defining $m^2\equiv\rho_L+\rho_R$ 
and $2\tilde \theta\equiv \theta_L-\theta_R$ puts the action (\ref{topsupercact}) into the form
\be
  S_{\rm TSC}^{(3+1)}=\int d^{3+1}x \left(-\frac{1}{4 e^2}\tilde F_{\mu\nu} \tilde F^{\mu\nu}+\frac{\tilde   
  \theta}{32\pi^2}\epsilon^{\mu\nu\rho\sigma}
  \tilde F_{\mu\nu}\tilde F_{\rho\sigma}+\frac{m^2}{2}\tilde A_\mu^2\right)\;.\ 
  \label{topsupercactred}
\ee
The tildes on $F$ and $A$ are added to emphasize that there is a transformation that relates them to the quantities in (\ref{topinsact}). Indeed, at suffienctly large energies where we can ignore the mass term for the photon $\tilde A_\mu$, this relation is nothing but the usual Maxwell duality, 
\be
  F_{\mu\nu}=\frac{1}{2}\epsilon_{\mu\nu\rho\sigma}\tilde F^{\rho\sigma}\;,
\ee
supplemented by the choice $\tilde \theta=\theta$. The duality can be extended to the level of the path integral by writing a first order master action, 
\be
S_{\rm master}^{(3+1)}=\int d^{3+1}x\left[\frac{1}{4 e^2}\left(2\epsilon^{\mu\nu\rho\sigma}F_{\mu\nu}\d_\rho A_\sigma-F_{\mu\nu}F^{\mu\nu}\right)
+\frac{\theta}{8\pi^2}\epsilon^{\mu\nu\rho\sigma}\d_\mu A_\nu \d_\rho A_\sigma\right].
\ee
If we vary it with respect to $A_\mu$, we obtain the constraint $\epsilon^{\mu\nu\rho\sigma}\d_\nu F_{\rho\sigma}=0$, which can be solved by 
\be
F_{\mu\nu}=\d_\mu A_\nu -\d_\nu A_\mu \equiv \d_\mu \tilde A_\nu-\d_\nu \tilde A_\mu\;,
\ee
where in the last equality we have denoted the $\tilde A_\mu$ field by the same name as the field we varied. The $\theta$-term is topological and so does not contribute to the equation of motion. Nevertheless we can, and should in fact, keep retain it in the action as it depends on the solution for $A_\mu$. Finally, we obtain
\be
S_{\rm TSC}^{(3+1)}
=\int d^{3+1}x\left(-\frac{1}{4e^2}(\d_\mu \tilde A_\nu -\d_\nu \tilde A_\mu)^2+\frac{\theta}{8\pi^2}\d_\mu \tilde A_\nu \d_\rho \tilde A_\sigma\right)\;,
\ee
which is nothing but (\ref{topsupercactred}) if we ignore the mass term. On the other hand, varying the master action with respect to $F_{\mu\nu}$, obtains the equation of motion 
\be
F_{\mu\nu}=\frac{1}{2}\epsilon_{\mu\nu\rho\sigma}(2\d^\rho A^\sigma)\;,
\ee
which, after substitution back into the action, gives 
\be
S=\int d^{3+1}x\left(-\frac{1}{4e^2}(\d_\mu A_\nu -\d_\nu A_\mu)^2+\frac{\theta}{8\pi^2}\d_\mu A_\nu \d_\rho A_\sigma\right) = S_{\rm TI}^{(3+1)}\;,
\ee
which is, of course, the action (\ref{topinsact}) for the topological insulator. Explicitly, the relation between the two fields is the usual Maxwell duality
\be
F_{\mu\nu}=\d_\mu \tilde A_\nu -\d_\nu \tilde A_\mu=\frac{1}{2}\epsilon_{\mu\nu\rho\sigma} 2\d^\rho A^\sigma.\label{maxdual}
\ee
This confirms our statement that in 3+1 dimensions, topological insulators and  topological superconductors are related through Maxwell electric-magnetic duality.

\section{Three dimensional topological superconductor - topological insulator duality}

Let's now drop down one spatial dimension and consider the 2+1 dimensional case. Here, if we rename the coefficient to $m/2$ and add a conventional Maxwell term, the action for the topological insulator (\ref{topinscs}) reads
\be
  S_{\rm TI}^{(2+1)}=\int d^{2+1}x\left[-\frac{1}{4e^2}F_{\mu\nu}^2-\frac{m}{2}\epsilon^{\mu\nu\rho}
  A_  
  \mu \d_\nu A_\rho\right].
\ee
This action is usually described as a ``topologically massive Maxwell theory". Defining
\be
  F^\mu\equiv \epsilon^{\mu\nu\rho}\d_\nu A_\rho\;,\label{fmudef}
\ee
it can be put into the form
\be
{S'}_{\rm TI}^{(2+1)}= \int d^{2+1}x\left[\frac{1}{2}F_\mu F^\mu- \frac{m}{2} F^\mu A_\mu\right].
\ee
On the other hand, for a topological superconductor, we would obtain a Chern-Simons term, 
which can be thought of as the dimensional reduction of the $\theta$-term as well as a mass term for the photon, in the same way as we saw in 3+1 dimensions. 
Consequently,
\be
  S_{\rm TSC}^{(2+1)}=\int d^{2+1}x\left[\frac{m}{2}\epsilon^{\mu\nu\rho}\tilde A_\mu \d_\nu \tilde 
  A_\rho-\frac{m^2}{2}\tilde A_\mu \tilde A^\mu\right].
\ee
This action was cited in \cite{Townsend:1983xs} as manifesting so-called ``self-duality in odd dimensions". Strictly speaking, we should also have added a 
Maxwell term. However, at sufficiently low energies\footnote{This should be contrasted with the high energy limit of the previous section, where we neglected the 
mass term.} where $m \gg 1$, that term will be subdominant with respect to the two terms above.\\

Again, we can write a master action for the duality between the actions $S_{\rm TI}^{(2+1)}$ and $S_{\rm TSC}^{(2+1)}$. 
This was actually already observed in \cite{Townsend:1983xs,Deser:1984kw}, to which we refer the reader for more of the technical details. Here, we will 
content ourselves to note that in the present context, the duality maps us between a topological insulator and a topological superconductor in three spacetime 
dimensions. The master action is
\be
  S_{\rm master}^{(2+1)}=\int d^{2+1}x\left[-\frac{1}{2}f_\mu f^\mu +f_\mu F^\mu -\frac{m}{2}
  F^\mu A_\mu\right]\;,
\ee
where as before $F_\mu$ is defined in (\ref{fmudef}). Eliminating $f_\mu$ through its equation of motion we arrive at the action ${S'}_{\rm TI}^{(2+1)}$ 
for the topological insulator. If instead we eliminate $A_\mu$ though its equation of motion, $A_\mu=\frac{f_\mu}{m}\equiv \tilde A_\mu$,
which we have renamed to $\tilde A_\mu$ in order to avoid confusion with the $A_\mu$ from ${S'}_{\rm TI}^{(2+1)}$, we get
\bea
  S_{\rm TSC}^{(2+1)}&=&\int d^{2+1}x\left[-\frac{1}{2}f^\mu f_\mu+\frac{1}{2m}\epsilon^{\mu\nu\rho}   
  f_\mu \d_\nu f_\rho\right]\cr
  &=& \int d^{2+1}x\left[-\frac{m^2}{2}\tilde A_\mu \tilde A^\mu+\frac{m}{2}\epsilon^{\mu\nu  
  \rho}\tilde A_\mu\d_\nu \tilde A_\rho\right]\;,
\eea
which is the action $S_{\rm TSC}^{(2+1)}$ for the topological superconductor! In this case, the duality relation is given by
\be
  f^\mu=\epsilon^{\mu\nu\rho}\d_\nu A_\rho=m\tilde A^\mu\;,
\ee
and we notice that it can be put in the form of a particle-vortex duality relation if $\tilde A_\mu$ is locally a ``pure gauge", (of course, there is no 
gauge invariance due to the mass term, hence the quotation marks)
{\it i.e.} if $\tilde A_\mu=\d_\mu\phi$, with $\phi$ is the scalar 
Poincar\'{e} dual to $A_\mu$ in 2+1 dimensions. In terms of $\phi$ however, the topological superconductor action becomes the trivial one for a free scalar,
\be
{S'}_{\rm TSC}^{(2+1)}=  \int d^{2+1}x\left[-\frac{1}{2}(\d_\mu\phi)^2\right].
\ee
We also note that the duality relation can be obtained from the usual Maxwell duality in 3+1 dimensions (\ref{maxdual}), if we set $A_4=0$ and dimensionally reduce by setting 
\be
  \tilde A_\sigma(\vec{x},x_4)=e^{mx_4}\tilde a_\sigma(\vec{x});\;\;\;\;\; A_  
  \mu(\vec{x},x_4)=e^{mx_4}a_\mu(\vec{x})\;.
\ee
Consequently, 
\be
\epsilon^{\mu\nu \rho 4}\d_\mu A_\nu=\d_4\tilde A_\sigma\Rightarrow \epsilon^{\mu\nu\rho}\d_\mu a_\nu=ma_\sigma.
\ee

\section{Son's conjecture from particle-vortex duality}

Now we come to the crux of our note. In a remarkably insightful work \cite{Son:2015xqa}, Son proposed a low-energy effective theory for the composite fermion describing the half-filled Landau level in a 2+1 dimensional Fermi liquid, {\it i.e.,} the fractional quantum Hall effect at $\nu=1/2$ on the Jain sequence. He also went further to suggest that this effective theory is equivalent to a 2+1 dimensional
Dirac fermion theory. The Dirac fermion lives on a brane in a 3+1 dimenional bulk, and interacts electromagnetically through the bulk, via an action
\be
  S=\int d^{2+1}x\left[ i\bar\psi\gamma^\mu(\d_\mu-iA_\mu)\psi\right] - 
  \frac{1}{4e^2}\int d^{3+1}x\, F_{\mu\nu}^2\,.
  \label{interface}
\ee
This is a fermion zero mode on a domain wall, representing the surface mode of a 3+1 dimensional topological insulator. We want to understand the ground state and low energy excitations of the system in finite magnetic field $B=F_{xy}$, and specifically for the half-filled Landau level. 

Son's proposed low-energy effective theory, conjectured to be dual to the above Dirac fermion theory, is given by the action
\be
  S_{\rm eff}=\int d^{2+1}x \left(i\bar\psi\gamma^\mu(\d_\mu+2ia_\mu)\psi+\frac{1}{2\pi}  
  \epsilon^{\mu\nu\rho}A_\mu \d_\nu a_\rho\right)
  -\frac{1}{4e^2}\int d^{3+1}x F_{\mu\nu}^2 + \ldots\;,\label{son}
\ee
where now $\psi$ is a Dirac quasiparticle of the {\it composite fermion} type, $a_\mu$ is an emergent gauge field and $A_\mu$ is an external electromagnetic field with field strength 
$F_{\mu\nu}=\d_\mu A_\nu-\d_\nu A_\mu$. Note that $\psi$ is {\em electrically neutral} and carries charge only with respect to the {\em emergent} gauge field $a_\mu$. This is the same as in the case of the standard Halperin-Lee-Read (HLR) Chern-Simons-fermion theory 
for a composite fermion on the $\nu=1/2$ state, where the composite fermion is also electrically neutral. In fact, Son demonstrated a simple relation to the HLR theory. 

However, Seiberg and Witten argued in subsection 6.3 of \cite{Seiberg:2016rsg} that the coupling 
of $\psi$ to $2a_\mu$ is not correct, since issues of 
Dirac quantization mean that it doesn't give the right anomaly. This should be replaced, they argued, 
by $2a_\mu+A_\mu$, generating an electromagnetic Chern-Simons term
\be
  \frac{1}{4\pi}\epsilon^{\mu\nu\rho}A_\mu\d_\nu A_\rho.
\ee
Note that in perturbation theory $2a_\mu=2a'_\mu+A_\mu$ is just a redefinition, but in the full theory it is not allowed. Putting this together, the low energy effective action for the composite fermion proposed by Seiberg and Witten is 
\bea
  S_{\rm eff}&=&\int d^{2+1}x \left(i\bar\psi\gamma^\mu(\d_\mu+2ia_\mu+iA_\mu)\psi-\frac{1}  
  {2\pi}\epsilon^{\mu\nu\rho}A_\mu \d_\nu a_\rho
  -\frac{1}{4\pi}\epsilon^{\mu\nu\rho}A_\mu \d_\nu A_\rho\right)\cr
  &&-\frac{1}{4e^2}\int d^{3+1}x F_{\mu\nu}^2+...\;,\label{sw}
\eea
and it is this theory that should be dual to the Dirac fermion action (\ref{interface}). One of the physical consequences of Son's conjecture is a relation between the electron conductivity $\sigma$ and the conductivity $\tilde \sigma$ of the composite fermion which in the relativistic case reads,
\bea
  \sigma_{xx}&=&\frac{1}{4}\frac{\tilde \sigma_{xx}}{\tilde \sigma_{xx}^2+\tilde \sigma_{xy}^2}\cr
  \sigma_{xy}&=&-\frac{1}{4}\frac{\tilde \sigma_{xy}}{\tilde\sigma_{xx}^2+\tilde\sigma_{xy}^2}.
\eea
These can be combined into a single relation by writing $\sigma=\sigma_{xy}+i\sigma_{xx}$, 
so that
\be
  \sigma=-\frac{1}{4(\tilde \sigma_{xy}+i\tilde\sigma_{xx})}=-\frac{1}{4\tilde \sigma}\;,
  \label{soncond}
\ee
or, more symmetrically, $2\sigma=-\frac{1}{2\tilde\sigma}$. 

We would like to argue that this relation is a consequence of particle-vortex duality. Our intuition for this stems from \cite{Burgess:2000kj} where, indeed, particle-vortex duality (expressed as a 
relation on path integrals) was shown to give rise to a relation between the conductivity, $\sigma$, due to particles and the one due to vortices, $\tilde \sigma$, of the form (and for general anyonic particles)
\be
  \tilde\sigma=\frac{\frac{\pi}{\theta}-\sigma}{1-\left(\frac{\theta}{\pi}+\frac{\pi}{\theta}\right)  
  \sigma}.
\ee
Specifically, for bosons, $\theta=0$, and we find that 
\be
  \tilde \sigma=-\frac{1}{\sigma}.
\ee
Clearly,  if we replace $\sigma\rightarrow 2\sigma$ and $\tilde \sigma\rightarrow 2\tilde \sigma$, we get the relation obtained in \cite{Son:2015xqa} for fermions. In fact, the factors of two are perfectly sensible since, we need to keep in mind that the particles and vortices are bosons, specifically {\it composite} bosonic scalars made up of two fermions. In the case of the 
(topological) superconductor, we have Cooper pairs made up of two fermions, and we can expect a similar situation to hold on the (topological) 
insulator side. Therefore the conductivity of the Cooper pairs should be twice that of the fermions, leading to the relation  (\ref{soncond}).

If the particle and vortex scalars are made up of two fermions, and the fermionic actions (interacting with electromagnetism) are 
(\ref{interface}) and (\ref{sw}), the corresponding boundary scalar actions for the composite scalars, must be 
\be
  S_{\rm TI}=\int d^{2+1}x \left[-\frac{1}{2}|(\d_\mu-iq A_\mu)\Phi|^2-V(|\Phi|^2)-\frac{1}{4}
  F_{\mu\nu}^2+...\right]\;,\label{higgs}
\ee
where $q=2$, for the composite scalar of two Dirac fermions, and for the low energy effective (boundary) action of the 
composite (and electromagnetically neutral) scalar of two composite fermions, 
\be
  S_{\rm TSC}=\int d^{2+1}x\left[-\frac{1}{2}|(\d_\mu+2ia_\mu)\tilde\Phi|^2-V(|\tilde \Phi|^2)+\frac{1}{2\pi}  
  \epsilon^{\mu\nu\rho}a_\mu\d_\nu A_\rho
  -\frac{1}{4}F_{\mu\nu}^2+...\right].\label{vortexhiggs}
\ee
The relation between these two actions is exactly the particle-vortex duality described in, for example, \cite{Lee:1989fw,Zee:2003mt}. We review it here
for complentess. Writing $\Phi=|\phi| e^{i\theta}$, and assuming that $|\phi|$ is fixed at a value $v$ that minimizes the potential $V$, we get the 
first order action,
\be
S_{\rm P}=\int d^{2+1}x\left[\frac{1}{2v^2}\xi_\mu^2-\xi_\mu (\d_\mu\theta -q A_\mu)\right]\,,
\ee
in terms of an auxiliary field $\xi_\mu$. To implement the particle-vortex duality, we split the phase of the complex scalar, $\theta$, into a ``smooth" part and a part which encodes the $2\pi$ monodromy obtained by encircling the vortex,
\be
\theta=\theta_{\rm smooth}+\theta_{\rm vortex}\;.
\ee
Then, integrating out $\theta_{\rm smooth}$, we obtain the constraint $\d_\mu \xi^\mu=0$. This is solved by writing $\xi^\mu$ as the curl of a vector field
\be
  \xi^\mu=\epsilon^{\mu\nu\rho}\d_\nu a_\rho\;.
\ee
On substituting into the action, we obtain
\bea
  S_{\rm P}&=&\int d^{2+1}x\left[-\frac{1}{4v^2}f_{\mu\nu}^2+\epsilon^{\mu\nu\rho}\d_\nu a_\rho(\d_  
  \mu\theta_{\rm vortex}-qA_\mu)\right]\cr
  &=&\int d^{2+1}x\left[-\frac{1}{4v^2}f_{\mu\nu}^2+2\pi\, a_\mu j^\mu_{\rm vortex}-
  A_\mu J^\mu \right]\;,
\eea
where the vortex current is
\be
  j^\mu_{\rm vortex}=\frac{1}{2\pi}\epsilon^{\mu\nu\rho}\d_\nu \d_\rho\theta_{\rm vortex}\;,
\ee
and the current is $J^\mu=q\epsilon^{\mu\nu\rho}\,\d_\nu a_\rho$. To complete the description, which is now in terms of vortices, 
coupled to the new gauge field $a_\mu$, 
as evidenced by the presence of the vortex current in the action, one needs to introduce another (vortex) scalar field $\tilde \Phi$ that 
couples directly to $a_\mu$. Finally then, the description in the particle-vortex dual theory is via the action
\be
  S_{\rm V}=\int d^{2+1}x\left[-\frac{1}{4v^2}f_{\mu\nu}^2-\frac{1}{2}|(\d_\mu-i2\pi a_\mu)\tilde   
  \Phi|^2-V(|\tilde \Phi|)-A_\mu(q\epsilon^{\mu\nu\rho}\d_\nu a_\rho)\right]\;,
\ee
which is nothing but the action (\ref{vortexhiggs}) after a rescaling of $a_\mu$ by $2\pi$. 
This establishes our claim then, that (\ref{higgs}) and (\ref{vortexhiggs}) are particle-vortex dual.

\section{Conclusions}

By now it is well-established that duality, the fact that two {\it a priori} different theories can encode the same physics, furnishes a powerful tool to understand both 
sides of the correspondence. While this statement is more or less obvious in holographic systems, it is also true in more the prosaic low-energy, condensed 
matter context. As an example, in this note we have analyzed the duality between topological insulators and topological superconductors in three and four 
spacetime dimensions, its origins in particle-vortex duality and some of its physical implications. We have seen that in four dimensions, the actions for topological 
insulators and superconductors can be related, via a path integral transformation that realizes Maxwell electric-magnetic duality. In three dimensions, the same 
duality between topological insulators and superconductors is obtained from a transformation of topologically massive Maxwell theory with self-duality in odd 
dimensions. This, in turn, we showed could be understood in terms of particle-vortex duality.\\

\noindent
Son's conjecture of an equivalence of the 2+1 dimensional massless Dirac fermion, which corresponds to the boundary state of a topological insulator, to a low 
energy effective theory for the $\nu=1/2$ Landau level in terms of a composite fermion was shown also to be the result of a particle-vortex duality. The physical 
consequence in terms of conductivities of the dual theories was shown to be equivalent to the relation obtained in \cite{Burgess:2000kj} from particle-vortex 
duality. Finally, we showed also that the bosonic actions for the composite scalars made up of the fermions were are particle-vortex dual pairs.\\

\noindent
Duality, specifically low-dimensional dualities like electric-magnetic and particle-vortex dualities in four and three dimensions respectively, furnish a powerful tool 
to understand topological quantum matter. 
Like any good field, many more questions remain. For example, it is known that three dimensional topological superconductors manifest a fully time-reversal 
symmetric and gapped surface in the presence of strong interactions and a special kind of topological order \cite{Fidkowski:2013jua}. These superconductors are 
indexed by an integer $\nu$. Curiously, when $\nu$ is an odd integer, the topological order must be non-abelian. It would be very interesting to understand these 
topological superconductors in the light of the recent developments in particle-vortex duality \cite{Murugan:2014sfa, Metlitski:2015eka, Murugan:2015boa, 
Murugan:2015fvz}. Clearly though, we have only just scratched the surface. The bulk remains to open to exploration.

\section*{Note Added}
Particle-vortex duality remains the Cinderella of dualities, largely ignored but really a hidden gem. However, as we were writing up this work, we became 
aware of at least two more forthcoming articles by Karch \& Tong \cite{Karch:2016sxi} and, independently, Seiberg, Senthil, Wang and 
Witten \cite{Seiberg:2016gmd} 
 with some overlap with this one. Perhaps the time has 
come for particle-vortex duality to shine. 

\section*{Acknowledgements}

The work of HN is supported in part by CNPq grant 301219/2010-9 and FAPESP grant 2014/18634-9. 
The work of JM is supported by the National Research Foundation (NRF) of South Africa under 
GUN 87667. JM is grateful for the kind hospitality of the IFT where this work was initiated and 
the very generous support of the School of Natural Sciences of the Institute for Advanced Study, 
where it was completed. We would like to thank David Tong and Edward Witten for encouraging us to complete this work as soon as we did.

\bibliography{topcmt}
\bibliographystyle{utphys}

\end{document}